\documentclass[draft]{agujournal2019}
\usepackage{url} 
\usepackage{lineno}
\usepackage[inline]{trackchanges} 
\usepackage{soul}
\draftfalse

\journalname{Geophysical Research Letters}

\begin{document}

\title{Interplay of Three-Dimensional Instabilities and Magnetic Reconnection in the Explosive Onset of Magnetospheric Substorms}

\authors{Samuel R. Totorica\affil{1,2,3,4}, Amitava Bhattacharjee\affil{1,2,3,5}}

\affiliation{1}{Princeton Center for Heliophysics, Princeton, New Jersey 08543, USA}
\affiliation{2}{Department of Astrophysical Sciences, Princeton University, Princeton, New Jersey 08544, USA}
\affiliation{3}{Princeton Plasma Physics Laboratory, Princeton University, Princeton, New Jersey 08540, USA}
\affiliation{4}{International Research Collaboration Center, National Institute of Natural Sciences, Tokyo 105-0001, Japan}
\affiliation{5}{Center for Computational Astrophysics, The Flatiron Institiute, New York, New York 10010, USA}

\correspondingauthor{Samuel R. Totorica}{totorica@princeton.edu}



\begin{keypoints}
\item For the first time, an exact kinetic magnetotail equilibrium is used to model magnetospheric substorm onset
\item Comparing 2D and 3D simulations reveals the importance of the coupling between magnetic reconnection and the kinetic ballooning instability
\item Self-consistent particle trajectories are analyzed for the first time in a realistic fully kinetic magnetotail configuration
\end{keypoints}

\begin{abstract}
Magnetospheric substorms are preceded by a slow growth phase of magnetic flux loading and current sheet thinning in the tail. Extensive datasets have provided evidence of the triggering of instabilities at substorm onset, including magnetic reconnection and ballooning instabilities. Using an exact kinetic magnetotail equilibrium we present particle-in-cell simulations which capture the explosive nature of substorms through a disruption of the dipolarization front by the ballooning instability. We use self-consistent particle tracking to determine the nonthermal particle acceleration mechanisms.
\end{abstract}

\section*{Plain Language Summary}
Magnetospheric substorms are events featuring bursty flows of magnetized plasma, highly energetic particles, and intense polar auroras. Substorms play a key role in the response of the magnetosphere to variations in the incoming solar wind. The Earth's magnetic field lines are like elastic strings, and when they snap charged particles can be accelerated to high energies.  Additionally, when there is enough plasma pressure pushing against the magnetic field, the magnetic field lines can develop an unstable oscillation known as a ``ballooning instability'' which is driven by the alignment of the plasma pressure gradient with magnetic field curvature. Using computer simulations that follow the trajectories of billions of particles in the Earth's magnetosphere and compute their self-consistent electromagnetic forces, we show the importance of the interplay between reconnection and ballooning in the onset of substorms and acceleration of charged particles to high energies. These results have strong implications for the development of accurate models to predict space weather events and mitigate their damaging effects on critical infrastructure.

\section{Introduction}

The Earth's magnetosphere is a complex environment involving the dynamics
of magnetized, collisionless plasmas. The interaction with the incoming
magnetized solar wind creates the interface regions of the bow shock and the
downstream magnetosheath, and stretches the Earth's magnetic field into an
extended magnetotail on the night side of Earth. During times of suitable
interplanetary magnetic field orientation, the solar wind plasma can enter
the magnetosphere through magnetic reconnection, with excess energy
and magnetic flux becoming stored in the magnetotail. This eventually leads
to explosive disruptions known as substorms, where the magnetotail becomes
unstable and violently releases the stored energy \cite{Sitnov2019}. 
Magnetospheric substorms can be frequent, occurring several times per day
and associated with transient features such as bursty flows of magnetized
plasma, nonthermal particle acceleration, and intensification of the polar
aurora \cite{Birn2012}. Substorms play a key role in the global evolution of
the magnetosphere and its response to the variable solar wind, and
understanding their dynamics is critical for the development of accurate
models for geomagnetic activity that can predict space weather events and mitigate
their destructive impacts.

Critical unsolved problem in magnetotail physics include understanding the
physical mechanisms that lead to the onset of substorms
at near-Earth distances ($\sim 10 R_{E}$, where $R_{E}$ denotes the radius
of the Earth) and the associated particle energization
that leads to auroral intensification.
Considering the magnetic field reversal across the magnetotail current sheet
at mid-tail distances ($> 30 R_{E}$) and the nature of observations such as intermittent bursts of plasma flows
with enhanced electromagnetic fields, magnetic reconnection is widely
believed to play an important role in the disruption of the magnetotail.
Past work investigating the possibility of the onset of the tearing
instability in the undisturbed magnetotail has faced difficulty due to the
stabilizing effect of even weak normal (perpendicular to the reconnecting
field in the reconnection plane, or N component in LMN current sheet coordinates) magnetic field components, which are
present in the curved field lines of the magnetotail
\cite{Schindler1974,Galeev1976,Pritchett1991,Wang1993,Pritchett1994}.
However, numerical simulations have shown that solar wind driving
can lead to current sheet thinning and a reduction of the normal field,
eventually surpassing a threshold that enables reconnection onset through
collisionless tearing \cite{Pritchett2005,Pritchett2010,Bessho2014,Liu2014}.

Free energy associated with the Earthward plasma pressure gradient indicates the additional possibility of three-dimensional cross-tail
instabilities. Of particular interest, the aligned plasma pressure
gradient and magnetic field curvature vectors can lead to the growth of the
ballooning instability, which modulates the magnetic field
orthogonal to the reconnection plane
with linear growth rates that peak around
wavelengths of $k_{\perp} \rho_{i} \sim \mathcal{O} (1)$
\cite{Roux1991,Cheng1998}. The ballooning instability could become
destabilized as the magnetic field curvature increases from current sheet
thinning, and may operate simultaneously with reconnection.  It is believed
that the inclusion of kinetic effects in models of ballooning is
critical for explaining observations in the magnetotail, and that the region
of highest instability is in the near-Earth dipole field transition region
\cite{Cheng1998,Cheng2004}.

The details of how reconnection and/or ballooning lead to
substorm onset and nonthermal particle acceleration in the collisionless plasmas of the type present
in the magnetotail, and the interplay between near-earth and
distant tail effects, are not yet fully understood.  
Observations give strong support for substorm onset
auroral signatures being associated with an azimuthally localized disruption
in the near-Earth dipole-tail transition region
\cite{Donovan2008,Sergeev2012}, and the properties of the near-Earth
disruption appear consistent with kinetic ballooning 
\cite{Panov2012,Panov2012a,Nishimura2016}. Evidence exists for events with
effects further in the tail (such as reconnection flows) preceding
\cite{Donovan2008} or following \cite{Nishimura2016} the near-Earth
disruption, indicating the
possibility of multiple classes of substorms and onset mechanisms.
Recently,
3D simulations are starting to be used to directly investigate
the interaction of reconnection and ballooning
\cite{Vapirev2013,Sitnov2014,Zhu2014,Lapenta2015,Zhu2017,Pritchett2017,Pritchett2018,Lu2018}.

Initial conditions for simulations of the magnetotail have been a
pervasive challenge due to the complex magnetic field geometry.  Generalized Harris sheets that contain a
small normal component of the magnetic field \cite{Bessho2014,Liu2014} are a suitable model for
the distant tail but neglect the critical near-Earth region. 
Magnetohydrodynamic simulations can numerically relax to
realistic field profiles \cite{Hesse1993}, however these are not Vlasov
equilibria that could be used for kinetic simulations. While fluid models
give important insight into global aspects of the magnetotail \cite{Bhattacharjee1998,Bhattacharjee1998a,Birn1997,Zhu2014}, features of
substorms firmly established by observations such as the kinetic scales of thin current sheets and the range of plasma beta and wavelengths over which instabilities are observed show
the importance of kinetic dynamics in
substorms.
Fully kinetic simulations with realistic magnetic field profiles have 
previously been performed using a two-step procedure of first evolving ions 
in the fields as test particles, and then adding electrons at the locations
of the ions \cite{Pritchett2018}.  
However, simulations that begin with initial conditions that are not rigorously in equilibrium have the potential to introduce transient flows that could influence 
the later development of the system. With a lack of rigorous initial 
conditions, highly idealized generalized Harris sheets remain the standard 
for kinetic magnetotail studies.

In this Letter, we investigate the onset of magnetospheric substorms
using \textit{ab initio} particle-in-cell (PIC) simulations \cite{Birdsall}.
Using the fully relativistic, state-of-the-art PIC code OSIRIS
\cite{Fonseca2002a,Fonseca2008,Fonseca2013,Hemker2015}, we perform fully
kinetic and electromagnetic two- (2D) and three-dimensional (3D)
simulations, for the first time starting from an \textit{exact} kinetic
equilibrium that captures both the near-Earth dipole magnetic field and the
tail current sheet. Solar wind driving leads to current-sheet
thinning and the onset of reconnection, producing dipolarization fronts
and plasmoids that are ejected from the reconnection layer.  Comparing 2D
and 3D simulations allows the study of the
complementary roles of
reconnection and intrinsically 3D cross-tail instabilities, and their
impacts on dipolarization fronts and substorm onset.  We demonstrate that only in 3D, dipolarization
fronts become disrupted as they
travel into the near-Earth region as a result of the ballooning instability.
Electrons and ions are accelerated to suprathermal energies and
field-aligned currents traveling to the ionosphere are produced.  These results reveal
the importance of the coupling between reconnection and ballooning in
the magnetotail and provide new insight into substorm onset.

\section{Simulations}

As emphasized above, a novel feature of our simulations is that we use as initial
conditions an exact kinetic equilibrium that captures both the near-Earth
dipole field and the extended tail current sheet. 
The equilibrium we employ was developed as an exact solution of the Grad-Shafranov
equation using boundary conditions that approximate observations of the
magnetotail field structure \cite{Manankova2003,Yoon2005}.
Free parameters of the model include a factor $\gamma$ that characterizes
the relative strength of the dipole and tail magnetic fields, and the
tail current sheet thickness $\delta$.  We choose a value of $\gamma=-3$ which
gives the closest approximation to the commonly used model of Tsyganenko
\cite{Tsyganenko1989} while also not including an \textit{X} point in the initial
configuration.  The initial current sheet thickness is chosen to be
$\delta / d_{i} = 10$ (where $d_{i}$ is the ion inertial length) which is in the range of typical values for the magnetotail and ensures initial
stability of the tail.
The number of ion inertial lengths per Earth
radii is set to be $R_{E} / d_{i} = 5$.
Along the $x$-axis, which points in the
anti-sunward direction, the simulation domain extends from
$x = 7 - 52 R_{E}$, and of particular importance captures
the dipole to tail transition region around $x \sim 10-12 R_{E}$.
The $z$-axis is aligned with the Earth's magnetic axis and has a length $L_{z} = 15 R_{E}$.  The simulations are run for $550 \Omega_{ci}^{-1}$.
For the 3D simulations, the total length of the out-of-plane $y$-dimension
ranges from
$L_{y} = 2.5-10.0 R_{E}$, comparable to the typical cross-tail width of bursty
bulk flows inferred from observations.
The initial temperature ratio is set to be $T_{i} / T_{e} = 5$, following
observations, and a background plasma is included with density $n_{bg} = 0.1$.
The resolution of the spatial grid is $\Delta x / d_{e} = 0.5$, where
$d_{e}$ is the electron inertial length, and the
number of particles per cell per species (electrons and ions) is
$n_{ppc} = 16 \; (128)$ in 3D (2D). 
To make the 3D simulations computationally feasible we use an artificial
mass ratio of $m_{e} / m_{i} = 1/25$ and an
Alfv\'{e}n speed of  $V_{A} / c = 0.1$,
where $V_{A}$ is calculated using the parameters of the tail current sheet.

To model the influence of solar wind driving, we apply
an external electric field at the $z$ boundaries that
the $z$ boundaries that drives plasma flows into the current layer \cite{Liu2014,Bhattacharjee1998,Bhattacharjee1998a}.
The field profile along the $z$ boundaries decays
monotonically from the Earthward to tailward boundaries
along $x$, and decays from the $z$ boundaries to zero
at $z=0$ along $x$ boundaries.  The functional form
defined only along the boundaries is given by $E_{y}(z,x) = \left | \textup{sin}(\pi z / L_{z}) \right | ( 0.5 + 0.5 \, \textup{tanh}(17 - 2.5 \, x/R_{E}))$
where $E_{0} / V_{A} B_{0} = 0.1$ and $V_{A}$ and
$B_{0}$ are the values corresponding to the lobe
field.
This field profile leads to reconnection onset at $x / R_{E} \approx 20$,
in accordance with typical observations.
The $y$ dimension is periodic and
the $x$ and $z$ boundaries are conducting for the
electromagnetic fields and reflecting for the particles.
In the absence of driving the simulations are confirmed to be stable.

\section{Results}

\begin{figure}[htp]
\begin{center}
\includegraphics[width=0.75\textwidth]{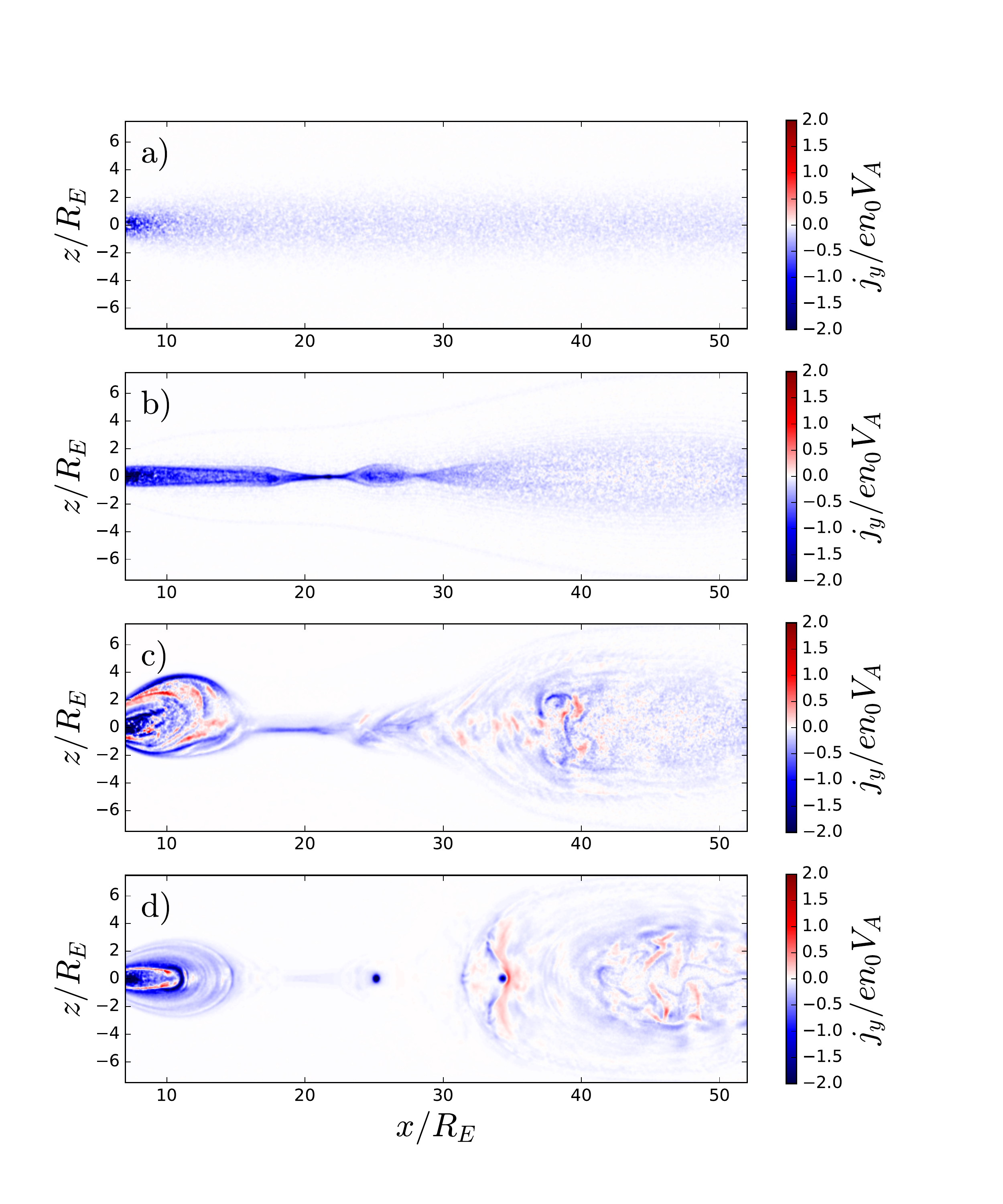}
\caption{\label{fig:3d_evolution} Cross-tail current density $J_{y}$
in a 2D slice in $z-x$ of the 3D domain at (a) $\Omega_{ci} t = 0$,
showing the initial equilibrium, (b) $\Omega_{ci} t = 449$ after
reconnection onset, showing plasmoids and the dipolarization
front propagating Earthwards, and (c) $\Omega_{ci} t = 520$ showing the
disruption of the dipolarization front and flapping of the
reconnecting current sheet. (d) shows a 2D simulation at $\Omega_{ci} t = 520$ for comparison.}
\end{center}
\end{figure}

The evolution of the 3D simulation is illustrated in Figure
\ref{fig:3d_evolution}, which shows a 2D slice of the
cross-tail current density in the $x-z$ plane at three different
times. 
The external electric field at the $z$ boundaries drives magnetic flux
and plasma flows towards the center of the current sheet at $z=0$, leading
to thinning of
the current sheet and the eventual onset of reconnection
once it reaches a scale on the order of $d_{e}$.
The thin current sheet where reconnection occurrs
shows growing oscillations in the $z$-direction, which
correspond to the current sheet
flapping motions that have been observed by satellites and
in simulations \cite{Sergeev2003,Runov2005,Petrukovich2006,Sharma2008,Runov2009,Sitnov2014,Wei2019}.  The characteristic length scale of the wavelength
and amplitude is seen to be on the order of $R_{E}$,
similar to observations \cite{Sergeev2003,Petrukovich2006,Runov2009}.
Reconnection results in both Earthward and tailward plasma
flows, and the formation of several plasmoids / flux ropes that are ejected both Earthward and tailward. After
the initial onset of reconnection there is an intense layer of
current that travels Earthward that is associated with a sharp
rise in the normal component of the magnetic field $B_{z}$ over
a length scale $\sim d_{i}$.
This corresponds to the dipolarization
fronts which are often seen in observations \cite{Angelopoulos1992}.
The layer becomes disturbed as it enters the near-Earth
dipole field region around $x / R_{E} = 10-12$ and
is eventually disrupted by a cross-tail instability.
In analogous 2D simulations in the $x-z$ plane where 
reconnection occurs without cross-tail instabilities,
the dipolarization front remains at the end of the
simulation.

\begin{figure}[htp]
\begin{center}
\includegraphics[width=0.75\textwidth]{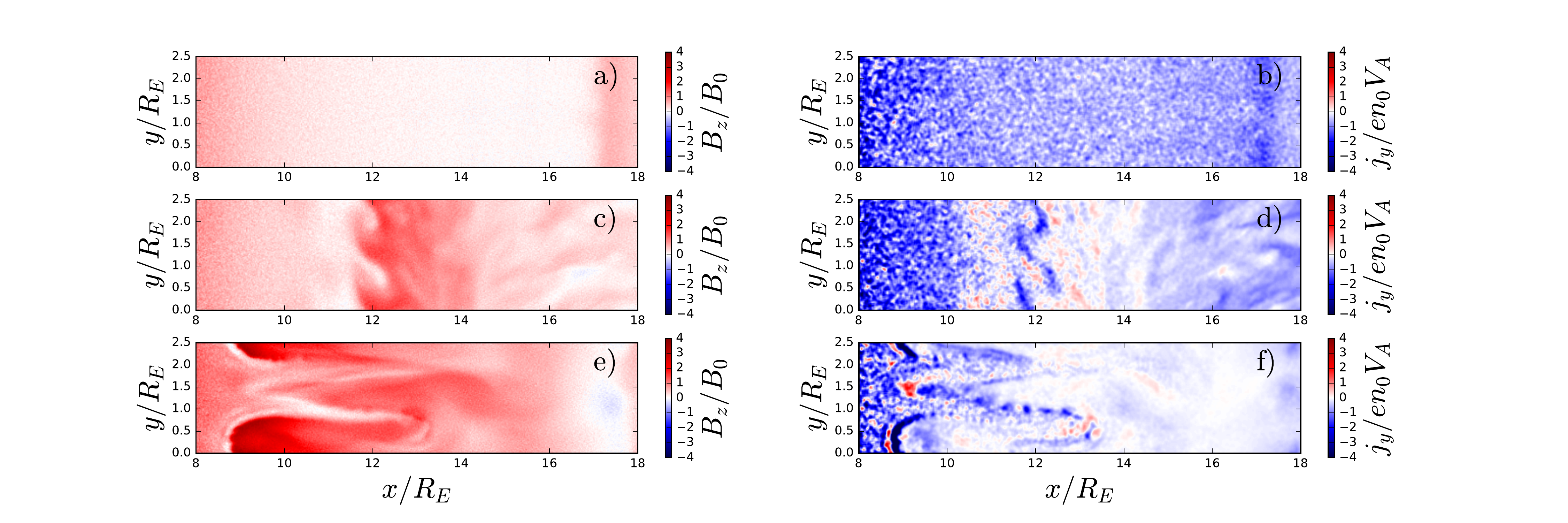}
\caption{\label{fig:equatorial} 2D slices in the equatorial $y-x$ plane
at $z=0$ of the normal
component of the magnetic field $B_{z}$ (a,c,e) 
and the cross-tail current density $J_{y}$ (b,d,f)
at three times.
(a,b) At $\Omega_{ci} t = 449$, the dipolarization front
propagates stably towards the Earth.
(c,d) At $\Omega_{ci} t = 484$, the dipolarization front
enters the near-Earth region and begins to
show modulations. (e,f) At $\Omega_{ci} t = 520$ the nonlinear
growth of the instability has resulted in a disruption of
the dipolarization front.}
\end{center}
\end{figure}

\begin{figure}[h]
\begin{center}
\includegraphics[width=0.95\textwidth]{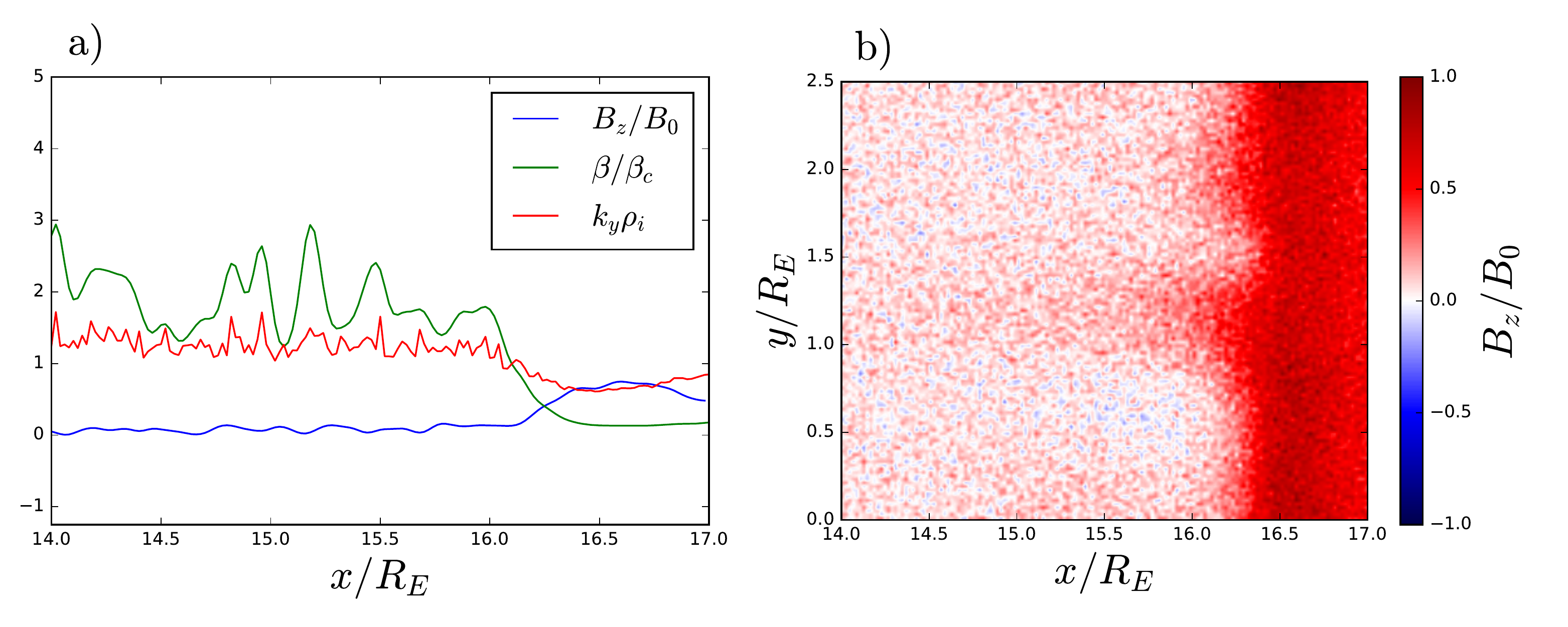}
\caption{\label{fig:gyro} a) Normal component of the magnetic field $B_{z}$
normalized to the characteristic field strength of the tail current sheet
$B_{0}$, plasma beta normalized to the critical value of $\beta_{c}=50$
for the onset of the kinetic ballooning instability, and $k_{y} \rho_{i}$ of
the instability in the near Earth region of the simulation.  b) $B_{z}$
component of the magnetic field in the near-Earth region of the equatorial
plane.  Analysis is performed at $\Omega_{ci} t = 456$.
}
\end{center}
\end{figure}

To understand the 3D mechanism of the dipolarization front disruption,
Figure \ref{fig:equatorial} shows the dynamics in the equatorial $x-y$
plane at $z=0$, showing the evolution of the normal component of the
magnetic field $B_{z}$ and the cross tail current density $J_{y}$.
The early time clearly shows the current layer of the dipolarization front
supporting the sharp increase in magnetic flux from the reconnection
outflows over a length scale of $\sim d_{i}$, consistent with observations.
The dipolarization front maintains its stability as it travels from the
reconnection layer through the distant tail-region.
Once it reaches the transition region where the Earth's dipole field becomes
significant, the dipolarization front begins to show growing modulations.
At later times the current layer associated with the dipolarization front is
seen to disrupt and become disconnected in the near-Earth region. The 
characteristic length scales
of the growing modes are $k_{y} \rho_{i} \approx 1$, which together
with the disruption in the near-Earth region and the associated
field-aligned currents (described below) indicates that this
is the result of the kinetic ballooning instability
\cite{Cheng1998,Cheng2004}.
Comparing to simulations with $L_{y}/R_{E}=5$ and $10$, the length scale of the dominant mode and onset location of $x \approx 15 R_{E}$ are found to be 
insensitive to the limited system size in the $y$-direction (Figure S1 in Supporting Information).  The $L_{y}/R_{E}=10$ simulation additionally 
develops a localized tailward breakout with a width of 
$\Delta_{y}/R_{E} \sim 4$, which may be related to reconnection onset 
occurring slightly further along the tail in this simulation and will be investigated in
further detail in future work.
As the dipolarization front and associated plasma and
magnetic field travel Earthward, the Earthward pressure
gradient builds up to $L_{P} = \left | P / \nabla P \right | \sim 0.5 R_{e}$.
For substorm conditions derived from observations including
$L_{P} \sim 0.5 R_{e}$ and $k_{y} \rho_{i} = 1$, the growth
rate of the kinetic ballooning instability was calculated
to be approximately $\omega_{KBI} = 0.1 s^{-1}$ or
$\omega_{ci} / \omega_{KBI} \approx 50$ using $\omega_{ci}$
for a magnetic field of B = 10 nT \cite{Cheng1998}. The
disruption of the
dipolarization front takes place over a timescale 
$\omega_{ci} t \sim 50$ (Figure \ref{fig:equatorial}) and is
thus consistent with a growth rate of the order of
magnitude of that expected for the kinetic ballooning instability.
Figure \ref{fig:gyro} shows how the disruption is occurring in a region where 
the plasma beta exceeds the critical value of $\beta_{c} = 50$, consistent 
with the kinetic ballooning  instability \cite{Cheng1998}.
This mechanism is distinct from the BICI mode \cite{Pritchett2010b,Pritchett2011,Pritchett2013} which has $k_{y} \rho_{i} \gg 1$ and requires a small plasma beta.
The strong magnetization of the electrons in the dipolarization front precludes spurious numerical effects from
the artificial mass ratio such as an artificial
ion-electron drift-kink mode \cite{Daughton1999a}.  A physical
ion-ion kink mode from the initially stationary
background plasma used in the simulation could
potentially contribute to the disruption depending on
the specific value chosen for the background plasma
density.

\begin{figure}[htp]
\begin{center}
\includegraphics[width=0.75\textwidth]{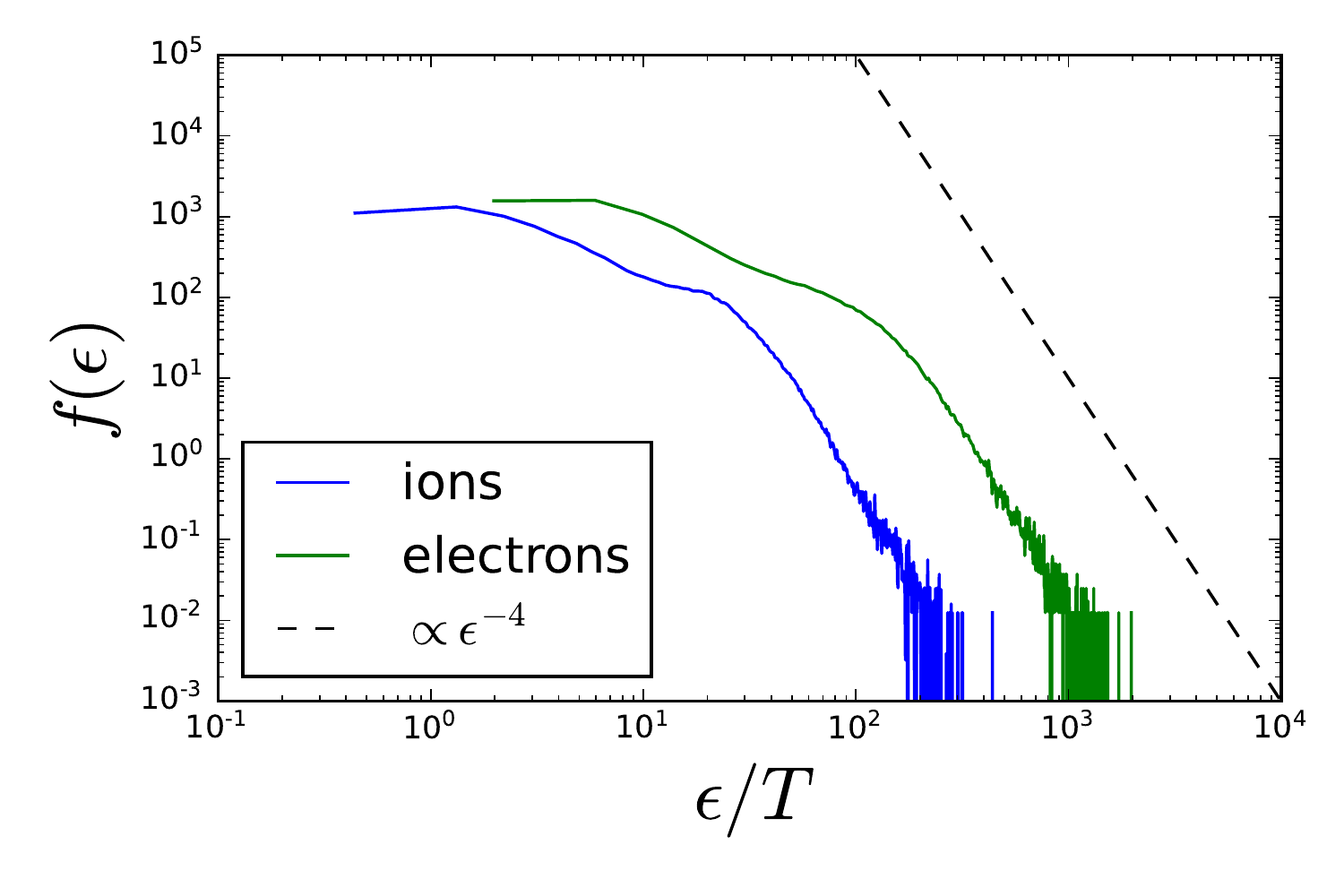}
\caption{\label{fig:spectra} Electron and ion energy spectra 
at $\Omega_{ci} t = 534$ integrated across $y$ within a square region of 1 ion skin depth width
centered at $x = 11\;R_{E}, z = 0.0\;R_{E}$. A 
power-law energy spectrum with index $p=-4$ plotted as reference.
}
\end{center}
\end{figure}

\begin{figure}[htp]
\begin{center}
\includegraphics[width=0.75\textwidth]{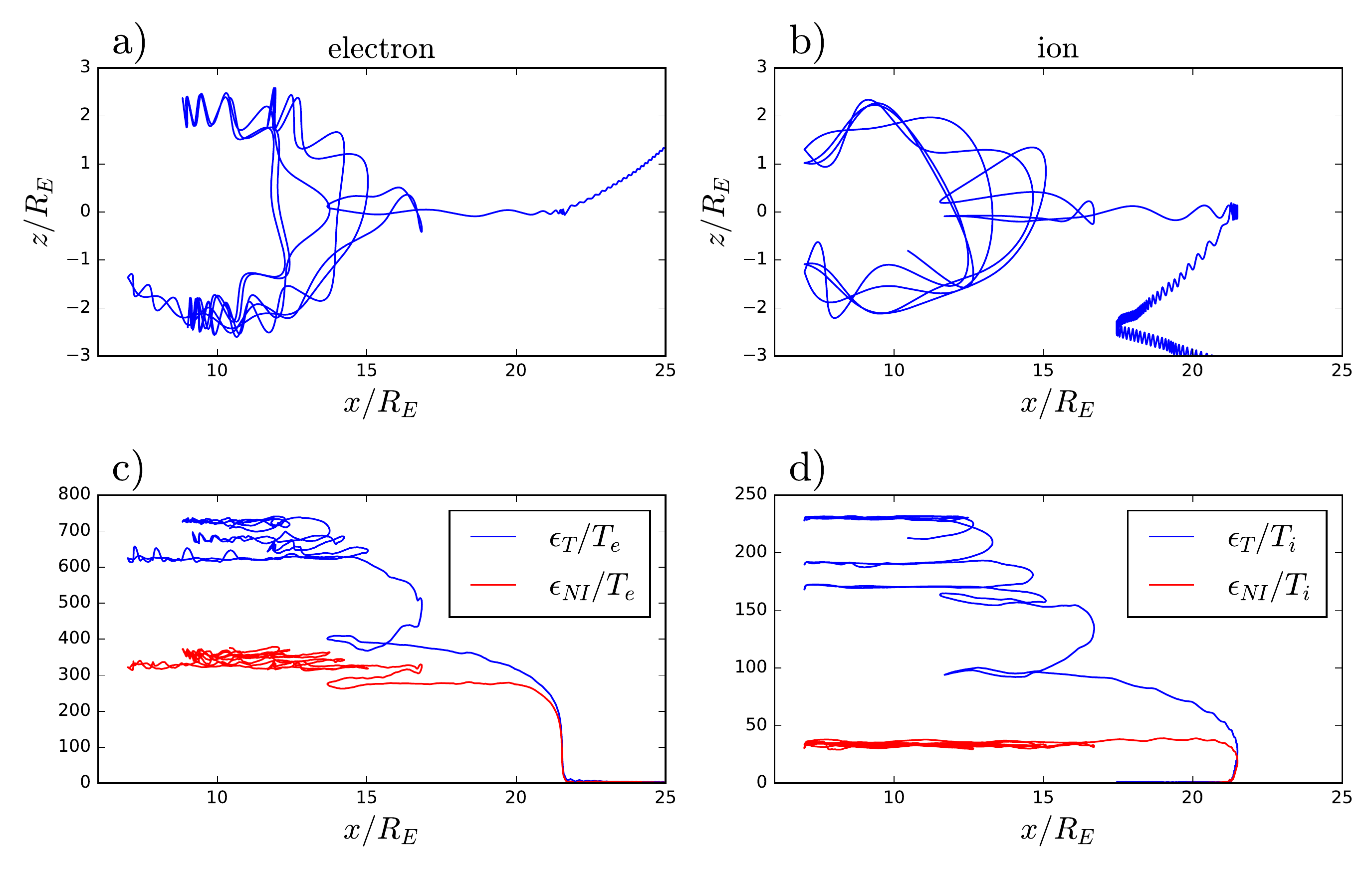}
\caption{\label{fig:tracking} Particle trajectories for
a representative energetic electron (a,c) and ion (b,d).
(a) and (b) show the trajectories in the $x$-$z$ plane and (c) and (d) show the total particle energies $\epsilon_{T}$ and the energy gained from the nonideal electric field $\epsilon_{NI}$ as a function of
position along $x$. The electron and ion first enter the plotted subdomain at $\Omega_{ci} t = 465$ and $\Omega_{ci} t = 165$, respectively, and the trajectories are plotted until $\Omega_{ci} t = 549$.}
\end{center}
\end{figure}

One of the most dramatic and readily observed signatures of
substorms is the intensification of the aurora that results
from charged particles impinging on the
ionosphere \cite{Birn2012}. The parallel component of the current
$J_{\parallel}$ shows Earthward traveling field-aligned currents originating
at the dipolarization front that intensify as the dipolarization front is
disrupted. These field-aligned currents travel to the Earthward boundary,
and in an open system could continue to travel and impact the ionosphere
and give rise to the auroral intensification characteristic of substorms.
Nonthermal energetic electrons and ions are also commonly observed in
association with substorm onset. Particles are known to be energized
both in the vicinity of the reconnection site \cite{Moebius1983,Oieroset2002,Imada2005,Imada2011,Imada2015a}
and in the near-Earth region \cite{Baker1981,Birn1997,Birn1997a} (referred
to as substorm injections). To investigate the particle acceleration
resulting from the disruption of the dipolarization front, Figure
\ref{fig:spectra} shows the electron and ion
energy spectra integrated over a square region of 1 ion skin depth width
centered at $x = 11\;R_{E}, z = 0.0\;R_{E}$.  As the dipolarization front crosses into the
near-Earth region and becomes disrupted, both electrons and ions show strong 
energization.  The spectra are normalized to the initial thermal energy
of each species, and the most energetic particles are seen to reach more
than 100 times the initial thermal energy.  In physical
units these electrons are reaching energies of
approximately $\sim 100 \textup{keV}$, and the highest energy
portion of the spectrum resembles a power law with an index
of $\sim 4$ over a limited energy range. This compares
favorably with
recent satellite measurements of energetic electron and ion acceleration
in the magnetotail \cite{Imada2015,Turner2016}.

To determine the physical mechanisms leading to the
acceleration of the energetic particles we have performed
self-consistent tracking of the trajectories of a sample
of the most energetic ions and electrons from the 2D simulation.
Figure \ref{fig:tracking} shows the evolution of a typical energetic
electron, with its trajectory in the $x-z$ plane in
Figure \ref{fig:tracking} (a) and its total energy
$\epsilon_{T}$ as function of $x$
in Figure \ref{fig:tracking} (c) (blue).
The electron first gains a significant amount of energy
from the reconnection electric field as it enters the
current sheet near the \textit{X} point.  It is then directed earthward by the magnetic field
and begins to travel
along the Earth's dipolar field lines and experience mirror
reflections as it approaches the polar regions.  The
electron continues to gain energy from betatron
acceleration as magnetic flux piles up from the
Earthward travelling reconnection flows.  Electrons that
reach the dipolar field lines with larger pitch angles
are confined closer to the equatorial plane where the cross
tail electric field resulting from the flux pileup is
greatest, and are accelerated at a greater rate.
Other trajectories also show electrons can become
trapped inside plasmoids and gain energy in a Fermi acceleration
process as they bounce between the ends of the plasmoid
\cite{Drake2006b} which are ejected and absorbed into the Earths dipole
field by a secondary reconnection processes.
The acceleration mechanisms for energetic ions are similar,
and a representative example is shown in
Figures \ref{fig:tracking} (b) and (d).  Ions are also
seen to gain energy from the reconnection electric field at
the \textit{X} point and inside plasmoids.  Additionally,
the large gyroradii of the ions allows them to be
reflected upstream into the reconnection flows
(Figure \ref{fig:tracking} (d)) and further gain energy from
the enhanced cross tail electric field in the magnetized
plasma flow.  The red lines in Figure \ref{fig:tracking} (b) and (d)
show the energy gained from the nonideal component of the electric field ${\bf E}_{NI}$,
calculated self-consistently within the simulation as ${\bf E}_{NI} ={\bf E} + {\bf V} \times {\bf B}$, where ${\bf V}$ is the plasma fluid velocity.  The initial
energy gain from the X points comes largely from the nonideal field, while the remaining
energy gain during mirror reflections comes almost entirely from the ideal field.  This
shows the importance of kinetic effects for modelling particle acceleration in the
magnetotail, but lends support to methods that couple global fluid models to
kinetic models in certain regions of the domain.
These results confirm major conclusions
of previous theoretical and test particle studies of
particle acceleration in the magnetotail \cite{Birn2012},
for the first time in a self-consistent fully kinetic
simulation.

\section{Conclusion}

In conclusion, we have implemented a novel configuration of an exact kinetic
equilibrium that captures the dipole to tail transition and reproduces the
salient features observed in the magnetotail. Solar wind driving leads to
thin current sheet formation and the onset of reconnection, resulting in
intermittent plasma flows, plasmoid formation and ejection, and current
sheet flapping.  Earthward traveling dipolarization fronts are formed in
both 2D and 3D simulations, but only in 3D the dipolarization front is
disrupted in the near-Earth region by the ballooning instability.
Field-aligned currents and nonthermal ions and electrons are accelerated, and particle tracking has determined the dominant acceleration mechanisms.
Future work will involve incorporating these dynamics into global
magnetosphere simulations for predictive space weather modelling.



%
%
%
%
%
%
%
%
%
%

\section*{Data availability statement}
Simulation data is hosted on the Zenodo
repository and can be accessed at
\linebreak
https://zenodo.org/record/8310560
(DOI 10.5281/zenodo.8310560) \cite{data}.

\acknowledgments
The authors acknowledge the OSIRIS Consortium, consisting
of UCLA and IST (Portugal) for the use of the OSIRIS 4.0 framework. S.T. was
supported by the NASA Jack Eddy Postdoctoral Fellowship and the Max-Planck
Princeton Center. Simulations were performed on Perlmutter (NERSC), Mira and Theta (ALCF) and Bluewaters (NCSA).

\bibliography{magnetotail.bib}

%
%
%
%
%

\end{document}


%
%


\title{Supporting Information for ``Interplay of Three-Dimensional Instabilities and Magnetic Reconnection in the Explosive Onset of Magnetospheric Substorms''}
%
%

%
%



\authors{Samuel R. Totorica\affil{1,2,3,4}, Amitava Bhattacharjee\affil{1,2,3}}

\affiliation{1}{Princeton Center for Heliophysics, Princeton, New Jersey 08543, USA}
\affiliation{2}{Department of Astrophysical Sciences, Princeton University, Princeton, New Jersey 08544, USA}
\affiliation{3}{Princeton Plasma Physics Laboratory, Princeton University, Princeton, New Jersey 08540, USA}
\affiliation{4}{International Research Collaboration Center, National Institute of Natural Sciences, Tokyo 105-0001, Japan}

%
%

%

\begin{article}

%
%

\noindent\textbf{Contents of this file}
\begin{enumerate}
\item Figure S1
\end{enumerate}

\end{article}
\clearpage

\begin{figure}[htp]
\begin{center}
\includegraphics[width=0.9\textwidth]{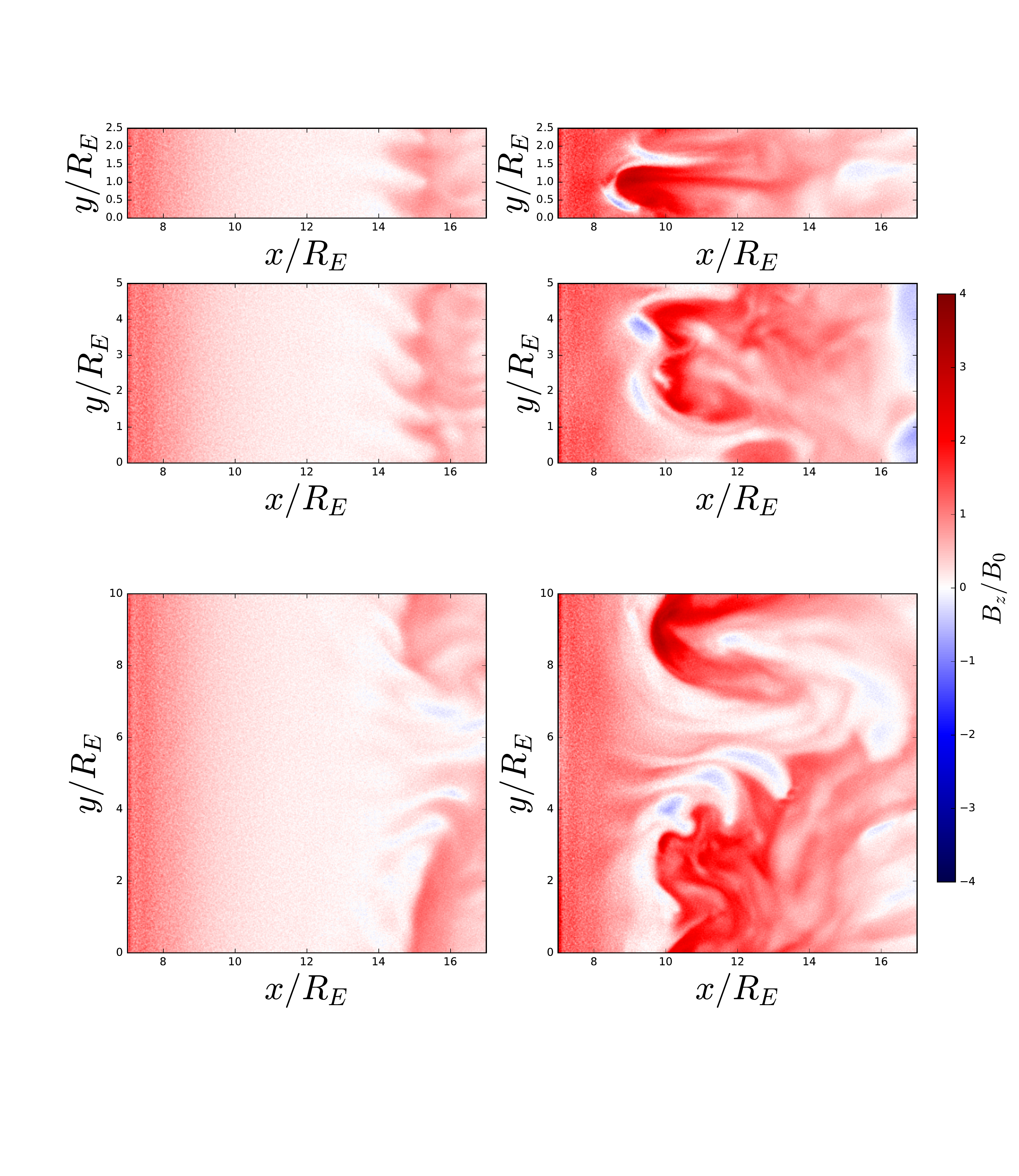}
\caption{Two-dimensional $x-y$ slices of the 
$B_{z}$ magnetic field component from simulations with three different system 
sizes in the $y$ direction: $L_{y} / R_{E} = 2.5$ (top), $L_{y} / R_{E} = 5.0$ (middle), and $L_{y} / R_{E} = 10.0$ (bottom). The different system sizes show
a well defined wavelength at early times that is well captured at all system 
sizes, and a qualitatively similar nonlinear phase with a strong distortion
of the dipolarization front, demonstrating that the dynamics in our 
simulations are not an artifact of the limited system size $L_{y}$.
The early and late times for the smaller two runs are $\Omega_{ci} t = 470$ and $\Omega_{ci} t = 513$, respectively.  For the $L_{y} / R_{E} = 10.0$ simulation,
reconnection onset occurs slightly further down tail, and for this case the
times of $\Omega_{ci} t =499$ and $\Omega_{ci} t =534$ are chosen to show
similar stages in the evolution for comparison.
}
\end{center}
\end{figure}